# The primary importance of the research question: Implications for understanding natural versus controlled direct effects and the "cross-world independence assumption"

Authors:
Ian Shrier, Centre for Clinical Epidemiology, Lady Davis Institute, Jewish General Hospital, McGill University, Montreal, Quebec, Canada

Etsuji Suzuki, Department of Epidemiology, Graduate School of Medicine, Dentistry and Pharmaceutical Sciences, Okayama University, Okayama Japan.

**Corresponding Author:**
Ian Shrier MD, PhD
Centre for Clinical Epidemiology, Lady Davis Institute, Jewish General Hospital,
3755 Cote Sainte Catherine Rd
Montreal, Quebec H3T 1E2, Canada
https://orcid.org/0000-0001-9914-3498
Email: ian.shrier@mcgill.ca
Phone Number: 1-514-229-0114

Abstract Word Count: 170
Manuscript Word Count: 3279

## Abstract
When developing new interventions to minimize the harmful effects and maximize the beneficial effects of an exposure, investigators usually target the mechanisms that mediate the causal effect of the exposure on the outcome. Predicting the causal effect of these new interventions is generally done through identifying either (1) the controlled direct effect, or (2) the pure (natural) direct effect. In this article, we use the interventionist approach to discuss how these two approaches answer different questions, and the additional underlying assumptions of each compared to the other. We use a specific example for the development of a new intervention that might reduce the harmful effects of smoking on chronic obstructive pulmonary disease by removing the inhalation of harmful chemicals.
## Keywords
Cross-world assumption, direct effects, causal inference, decomposition analysis

## Declarations
**Funding:** This work was unfunded
**Conflicts of interest/Competing interests**: The authors have no conflicts of interest to declare
**Availability of data and material**: not applicable
**Code availability**: not applicable
**Authors' contributions:** Both authors contributed to the writing of this manuscript.
**Ethics approval:** not applicable
**Consent to participate:** not applicable
**Consent for publication**: not applicable

# Commentary

The pure effects described by Robins and Greenland,(1) and later called natural effects by Pearl,(2) have been criticized because a cross-world independence assumption is often (but not always) required to identify them.(3, 4) The traditional cross-world independence assumption is usually stated within the context of an idealized simple randomized experiment where the controlled direct effect (CDE) can be estimated but the natural (or pure) direct effect (NDE) cannot.(5) However, other experimental designs are possible such as an idealized cross-over randomized trial where the participant returns to their baseline state after one arm of the trial.(6, 7) In fact, the cross-over randomized trial is more analogous to the potential outcomes framework. In this framework both CDE and NDE require measurements associated with three and only three single interventions (which may or may not be part of a single complex intervention in practice). We approach the problem from the spirit of John Tukey's "an approximate answer to the right question is worth far more than a precise answer to the wrong one".

Similar to recommendations by others, our approach requires clearly stating the research question, and just as importantly, clearly describing the actual interventions that would be necessary to answer the question.(5, 7) Our conceptual perspective is consistent with more recent work on mediation of natural effects.(7-11)

We use a simple example to estimate the causal effects of smoking on lung disease (Figure 1A), using only binary variables. Smoking (A) causes inhalation of dangerous chemicals (M), which causes an increased risk of chronic obstructive pulmonary disease (COPD) (Y). Occupation (e.g. firefighters) (C) causes inhalation of harmful chemicals leading to COPD, and there may also be a direct effect of heat-related COPD. Thus, C is a mediator-outcome confounder. Finally, L denotes every other cause of inhalation of harmful chemicals (e.g. air pollution, camp fires, and so on).

Within our "intervention framework", we could estimate the total causal effect of smoking on COPD by randomizing participants to smoke, or not to smoke. Although we have strong beliefs that inhaling harmful chemicals is creating most of the damage, we remain quite unsure about other ways smoking might cause COPD. Therefore, we might be very interested in obtaining the causal effect of smoking if we could somehow remove these harmful chemicals (known as the direct effect of smoking independent of the mediator "inhaled chemicals"). The idea that removing these chemicals might make smoking less harmful is what eventually led to "vaping", which is smoking without inhalation of many of the known harmful chemicals in smoke.

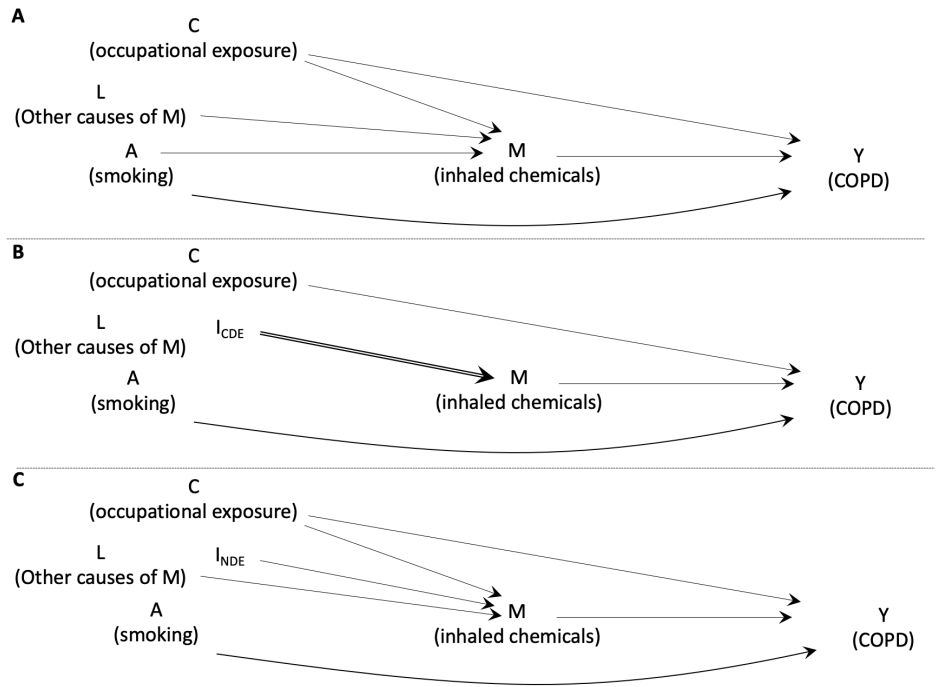

**Fig. 1** The causal diagram in panel A represents the naturally occurring context for the effects of smoking (A) on COPD (Y) acting partially through the mediator (M) of inhaled harmful chemicals. Occupation (C) may also cause inhaled chemicals, as well as COPD. In panel B, the causal diagram represents the context where one is interested in CDE. In this world, investigators fix M to a specific value (e.g., M = 0), which means they apply an intervention ($I_{CDE}$) that becomes the sole cause of M (deterministic, indicated by a double-arrow(12)), and all other arrows into M are removed. In panel C, the causal diagram represents the context where one is interested in NDE. In this world, investigators apply an intervention ($I_{NDE}$) that removes the effect of A on M (no arrow from A to M) but leaves all other causes affecting M intact. From Rothman's sufficient causal set framework, the removal of at least one component cause of each sufficient causal set that includes A (i.e. either A = 0 or A = 1) is sufficient for the $I_{NDE}$.

We highlight that direct effects are usually estimated by measuring the mediator in a study that is designed to estimate the total causal effect of the main exposure. However, if one were able to manipulate the mediator, one could answer these questions directly with a new four-arm randomized controlled trial (RCT) that evaluates the effects of simultaneously intervening on both smoking and the mediator.(7) In our example, the direct effect of smoking might be due to heat-related damage. Although we consider smoking as a single intervention for simplicity, it occurs over long periods of time and all our variables can be conceptualized as time-dependent.

## Controlled Direct Effects (CDEs)

The CDE in our example is the effect of smoking (A in Figure 1) that is not mediated through an increase in harmful chemical inhalation. Mathematically, $CDE_{M=m}$ is the effect that would be observed if everyone was switched from A = 0 to A = 1, while keeping M fixed at a particular value m for everyone (e.g. M = 0, one-world).

One challenge with some of the current literature on $CDE_{M=m}$ is that, even though authors carefully explain the intervention on the exposure of interest, authors rarely define what they mean when they "fix M" to the same value m for every person.(13) In an RCT (we will assume 100% adherence for simplicity), the total causal effect of smoking is obtained with an intervention that leads to smoking (A = 1) in the treatment group, and an intervention that leads to not smoking in the control group. However, hypothetical interventions on M to obtain the $CDE_{M=m}$ (i.e. $I_{CDE}$) also need to be specified. The $I_{CDE}$ must include a component that eliminates the influence of each of the following causal effects for inhalation of harmful chemicals: (1) smoking, (2) occupational exposure, and (3) every other cause of inhalation of harmful chemicals. In other words, $I_{CDE}$ must "overrule" the effects of every other variable that normally increases or decreases M (i.e. $I_{CDE}$ is deterministic for M). One possible example to set M = 0 for every person in the population is a complex intervention that includes highly effective filters at the end of cigarettes, prevention of occupational exposure to harmful chemicals, elimination of air pollution, ban camp fires, and so on. Alternatively, one "non-complex" intervention could be to implant a device into the trachea that filters all harmful chemicals. Essentially, this means that in the world with the new intervention present, M is no longer a function (i.e. caused by) of A, C, or any other variable other than $I_{CDE}$. Figure 1B illustrates this new world where $I_{CDE}$ is included (the double arrow from $I_{CDE}$ to M indicates deterministic relationship(12)), and all other arrows into M are removed. Although $I_{CDE}$ does not have to be generally included when drawing causal directed acyclic graphs (DAGs) because it is not a common cause of any two variables,(14, 15) it helps illustrate under-recognized assumptions in some contexts.

Within an intervention framework, investigators "fix" M to a value m with an intervention, and then simply randomize participants to A = 0 or A = 1 to obtain the $CDE_{M=m}$. If there is an interaction between the effects of A and M on Y, $CDE_{M=1}$, (i.e. setting M=1, where the complex intervention is smoking plus an intervention that causes all participants to inhale harmful chemicals) will be, by definition, different from $CDE_{M=0}$. The CDE calculated by including M in a regression model is actually a weighted average of these two different CDEs. However, we can also think about the concepts from a potential outcomes approach or cross-over trial with a complete return to baseline between interventions.

1. Fix M = 0 through the administration of $I_{CDE}$ (harmful chemical inhalation = 0) for every person throughout the experiment. No measurements.
2. Fix A = 0 (no smoking), measure Y.
3. Fix A = 1 (smoking), measure Y.

The $CDE_{M=0}$ is a contrast between the risk of COPD (Y) with smoking (intervention #3) and risk of COPD without smoking (intervention #2).

There are five important points to emphasize. First, as mentioned above, "fixing a variable" means we assume our theoretical $I_{CDE}$ will be deterministic for the value of M. Second, because we assumed A does not cause M in this world (Figure 1B), fixing either A first or M first leads to the same distribution of outcomes, as long as we measure Y after both variables are fixed. Here we assume that the fixed M does not change when A is subsequently intervened. Third, there is an assumption setting A = 0 does not affect the potential outcome of Y when A = 1, i.e. the participant's states before step 2 and before step 3 are identical ("washout" occurs). Fourth, the difference between the total effect and the CDE cannot in general be interpreted as an indirect effect. If there are no interactions between the effects of A and M on Y, however, $CDE_{M=1}$ and $CDE_{M=0}$ become identical, both of which are equivalent to NDE to be discussed in the next section.(16-18) Finally, even if A does not cause M, CDEs may differ from total effect if there is an interaction between the effects of A and M on Y.(17-19)

## Natural Direct Effects (NDEs)

The NDE is also the effect that occurs due to smoking that is not mediated through harmful chemical inhalation. However, it differs from the CDE because M is expected to be fixed by a complex intervention similar to $I_{CDE}$ but at different values for different participants. Mathematically, NDE is the effect that would be observed if everyone was switched from A = 0 to A = 1, while keeping M at the value it would have if A = 0 (noted in a potential outcome as M = M(0)). In this world, we assume the intervention to estimate NDE ($I_{NDE}$) leads to every individual's value of M being unaffected by their value of A. Therefore, the arrow from A to M is removed, but other factors such as environmental exposure, continue to affect M (Figure 1C).

In an RCT, our intervention to smoke (A = 1) or not smoke (A = 0) remains the same as our CDE example. However, to answer the NDE question, we only need to "remove" the causal effect of smoking on inhalation of harmful chemicals; we no longer require the intervention $I_{CDE}$ which retains M = m irrespective of occupational exposure or other causes on inhaling harmful chemicals. An $I_{NDE}$ in this context might to be the addition of a highly effective filter to the end of cigarettes (removing inhalation of harmful chemicals from smoke). More practically, one might consider vaping as a form of this $I_{NDE}$, even though it is not exactly the same. In any case, like the CDE, authors should specify how to intervene on M.

Now that we described the intervention, the mathematical approach using observed data requires fixing the value of M when A = 1 (one world) to what it would have been had A = 0 (a second world).(3) This is one reason for the "cross-world" terminology; we cannot obtain all relevant information from a single randomized experiment but rather, we require information from two different worlds. Within the potential outcomes approach, the cross-world independence assumption is expressed with the following notation:

$$Y(a, m) \perp M(a') \mid C$$

where Y(a,m) denotes a potential outcome of Y when A = a and M = m and M(a') denotes a potential outcome of M when A = a'. This has led some to question the meaningfulness of NDE, and also to the development of "interventional natural effects" that can theoretically be assessed through randomized trials.(20-22)

From the potential outcomes approach, like the CDE, we require only three interventions for the NDE, but in a different order. We again assume that the participant's states before step 2 and before step 3 are identical:

1. Fix A = 0 (no smoking), measure M (harmful chemical inhalation) and measure Y (COPD).
2. Fix M = the value after our first intervention (no smoking) for each participant through the administration of $I_{NDE}$. No measurements.
3. Fix A = 1 (smoking), measure Y (COPD).

The NDE is a contrast between the risk of COPD with smoking (intervention #3) and the risk of COPD without smoking (intervention #1).

As with CDE, there is no causal effect of A on M (Figure 1C) and therefore we can fix either A or M first. Viewed from this perspective, both the CDE and NDE can be assessed with three interventions and two measurements.

## Estimating CDE and NDE: Models, Assumptions and Definitions

Although we focused on the meaning of CDE and NDE, we briefly discuss estimation. First, although causal DAGs outline our hypothesized causal relationships in general, they do not identify all assumptions required for causal inference. Two models that are commonly used to display assumptions are a non-parametric structural equation model (NPSEM) and a finest fully randomized causally interpreted structural tree graph (FFRCISTG). A full discussion of the differences and implications of these models can be found in Robins et al.(7) In brief, both models require that all common causes of variables are included in causal DAGs; other variables can also be included if appropriate for the research question. The main difference is that an NPSEM assumes that the counterfactuals between variables have independent errors (sometimes referred to as NPSEM-IE(7)), and an FFRCISTG does not. Although a data generating process that violates the counterfactual independent error assumption is not common,(7) some examples have recently been provided.(7, 23)

Because the dependent errors noted above occur across counterfactuals that cannot be observed at the same time in a single randomized trial, this is called a cross-world assumption.(7) When the errors are dependent, the NPSEM associated with the causal DAG is not consistent with the data generating process, and inferences based on NPSEM-IE rules will sometimes be incorrect unless additional assumptions are made and are valid. However, even without assuming independent errors, bounds for the NDE can be estimated using the FFRCISTG model.(7) This means we can still obtain an approximate answer to our research question. Readers should be aware that it is possible to indicate dependent errors on the causal graph with a bidirectional arrow between the variables (graph now called acyclic directed mixed graph), or by including deterministic nodes for each counterfactual, but there remain challenges with these appproaches.(7)

The most common reported difference between identifying the NDE and CDE is that there is a necessary additional assumption that A is not a cause of a confounder of M-Y to identify NDE. This assumption would be violated if Figure 1A included an additional arrow from A to C. In this case, the confounder C is called a recanting witness because there is a path through C that is part of the direct effect (A → C → Y), and a path through C that is part of the indirect effect (A → C → M → Y).(24) If a recanting witness is present, the $CDE_{M=0}$ (combined effect through paths A → Y and A → C → Y) can be estimated because we are fixing M = 0 and therefore all the arrows going into M are removed. For the NDE, after removing the arrow A → M, A continues to affect Y through both the direct paths (A → Y and A → C → Y) and indirect path (A → C → M → Y). Therefore, NDE can only be estimated in very specific contexts.(4)

We now return to what fixing M means. From the interventionist framework, $CDE_{M=0}$ does not require the NDE assumption that A has no effect on C. However, it does require the assumption that one is able to remove two or more (depending on L) additional causal effects on M with $I_{CDE}$ compared to $I_{NDE}$.

Finally, there are two nuances about estimating direct effects. First, estimating both CDE and NDE are often said to require no unobserved confounding of the M-Y relationship.(7, 25) However, it is sometimes possible to estimate both CDE and NDE in this context, although estimating NDE still requires the independent error assumption noted above.(7) Second, in contrast to the common understanding that the absence of a "recanting witness" is a necessary condition to identify NDE, NDE can be estimated if the effect of exposure monotonically affects the confounder, or there is no additive interaction between the mediator and the confounder.(24) Without these additional restrictions, alternative methods can still estimate other types of direct and indirect effects of interest even if they do not represent NDE.(20)

**Clinical Relevance of CDE versus NDE**

We believe both CDE and NDE have important strengths and limitations. $CDE_{M=m}$ might be easier to estimate but it requires being able to describe all the interventions that will fix M = m for every participant. However, we should carefully consider whether it is meaningful to fix the amount of inhaled harmful chemicals for every participant from every cause to the same value, 0 or otherwise. In addition, when the exposure interacts with the mediator, including M as a covariate in a regression model yields a weighted average of $CDE_{M=0}$ and $CDE_{M=1}$;(16-18) the NDE naturally incorporates such interactions and the total effect is the sum of the natural direct and indirect effects. The underlying assumption of the intervention associated with $CDE_{M=0}$ is that one can eliminate all causes of harmful chemical inhalation (M), which seems daunting in epidemiology and public health. Conversely, the NDE informs on our causal question related to an intervention that only removes effects of smoking on chemical inhalation.

Despite our argument above, it may be easier to describe $I_{CDE}$ compared to $I_{NDE}$ in some contexts. For example, consider that ingestion of sugar drinks during running improves endurance time (26) and we are interested in knowing how much of the effect was mediated by the subsequent changes in insulin. One relatively simple $I_{CDE}$ might be to establish an intravenous line that measures insulin and simultaneously infuses or withdraws insulin from the blood depending on its concentration. This allows us to "fix" insulin without having to control each of the individual factors that can affect insulin secretion (e.g. hormones) or elimination. However, the $I_{NDE}$ would require blocking the release of insulin specifically due to sugar absorption from the gut, and still allow insulin to rise or fall for other reasons. Practically, the $I_{NDE}$ is a much more difficult intervention in this context.

We believe NDE is also helpful when deciding targets for new interventions. Our motivating example was to remove harmful chemicals from smoking, which was likely considered impossible 20 years ago. However, we may be able to achieve this through vaping of safe chemicals (notwithstanding cases under investigation(27)). Smoking in the presence of the intervention (i.e. without inhaling harmful chemicals such as vaping) may still have negative effects on COPD independent of M through heat-damage. Because the CDE estimates the effect when all harmful chemical inhalation is removed and not just those due to smoking, it is the NDE that addresses our question and would be considered of greater clinical relevance.

**General Limitations of CDE and NDE**
Although our objective did not include methods to estimate CDE and NDE, we emphasize that estimating these effects is a theoretical exercise with important challenges.(3, 4, 18) For example, whatever new intervention is developed to remove the effect of A on M is unlikely to be an idealized intervention that has one and only one effect.(5) The value of the new intervention will depend on the total causal effect across all paths, some of which may be unknown. The recent number of cases with lung damage of unknown causes from vaping (which may or may not eventually lead to COPD in the future) underscore the importance of not relying solely on estimates from any decomposition analysis, whether it is CDE or NDE.

As health researchers, we have to decide which of the different hypothetically beneficial interventions that eliminate particular causal effects should be tested given limited funding opportunities. We believe this requires we create a new causal diagram for each of the interventions. Each causal diagram would incorporate our knowledge about the direct and indirect paths in the presence of the intervention. One then simply chooses to test the intervention that *is expected* to lead to the greatest impact based on our synthesis of evidence and assumptions.

**Summary**
From the intervention perspective, estimating direct effects (both CDE and NDE) requires assumptions about the presence of a confounder of M and Y (including cross-world independence assumptions with or without a recanting witness). The NDE is meaningful if one is interested in developing an intervention that removes a specific cause of M, i.e. all effects of A on M. CDE is meaningful if one is interested in developing an intervention that removes every single cause of M. When NDE and/or CDE are not strictly identifiable from the data, there may be contexts where it is best to (1) estimate bounds for the effect, (2) consider a potentially biased answer to our question, or (3) consider that substituting a different question for the one originally posed provides the most valuable information possible.

# Supplementary Material
## Cross-World Independence Assumption
The traditional cross-world independence assumption is usually stated within the context of an idealized simple randomized experiment.(5) NDE is the effect that would be observed if everyone was switched from A = 0 to A = 1, while keeping M at the value it would have if A = 0 (noted in a potential outcome as M = M(0)). In a simple RCT, this information is not available because the world in which A = 1 and M = M(0) is never observed. Within the potential outcomes approach, the cross-world independence assumption is expressed with the following notation:

$$Y(a, m) \perp M(a') \mid C$$

where Y(a,m) denotes a potential outcome of Y when A = a and M = m and M(a') denotes a potential outcome of M when A = a'.

Because all causal analyses require assumptions, the requirement for one apparent additional assumption to estimate NDE is often mentioned as a reason to prefer CDE. Another commonly used reason to prefer CDE is related to the philosophical viewpoint that a "scientific question" requires that it is theoretically testable through experimental studies. If we limit ourselves to simple RCTs intervening on the exposure of interest, the NDE does not represent a scientific question. However, other experimental designs are possible and we could answer the question through an idealized cross-over randomized trial where the participant returns to their baseline state after one arm of the trial.(6, 7) In fact, the cross-over randomized trial is more analogous to the potential outcomes framework. From the crossover trial perspective, both CDE and NDE can be estimated with measurements associated with three and only three single interventions (which may or may not be part of a single complex intervention in practice).

## NPSEM versus FFRCISTG
Although causal DAGs outline hypothesized causal relationships in general, they do not identify all assumptions required for causal inference. Two models underlying causal DAGs that are commonly used are the non-parametric structural equation model (NPSEM) and the finest fully randomized causally interpreted structural tree graph (FFRCISTG). A full discussion of the differences and implications of these models can be found in Robins et al.(7)

In brief, both models require that all common causes of variables are included in causal DAGs; other variables can also be included if appropriate for the research question. The main difference is that an NPSEM assumes that the counterfactuals between variables have independent errors (sometimes referred to as NPSEM-IE(7)), and an FFRCISTG does not. Although a data generating process that violates the counterfactual independent error assumption is not common,(7) some examples have recently been provided.(7, 23)

Because the dependent errors noted above occur across counterfactuals that cannot be observed at the same time in a single randomized trial, this is called a cross-world assumption.(7) When the errors are dependent, the NPSEM associated with the causal DAG is not consistent with the data generating process, and inferences based on NPSEM-IE rules will sometimes be incorrect unless additional assumptions are made and are valid. However, even without assuming independent errors, bounds for the NDE can be estimated using the FFRCISTG model.(7) This means we can still obtain an approximate answer to our research question. Readers should be aware that it is possible to indicate dependent errors on the causal graph with a bidirectional arrow between the variables (graph now called acyclic directed mixed graph), or by including deterministic nodes for each counterfactual, but there remain challenges with both these approaches.(7)